\documentclass{article}

\usepackage{cite}
\usepackage{graphicx} 
\usepackage{lineno}
\usepackage[a4paper,top=3cm,bottom=2cm,left=2.5cm,right=2.5cm,marginparwidth=1.75cm]{geometry}

\usepackage[utf8]{inputenc}
\usepackage[T1]{fontenc}

\usepackage{xcolor}
\usepackage{graphicx} 
\usepackage{subfigure}
\usepackage{longtable}
\usepackage{multirow}
\usepackage{textcomp}
\usepackage{units}
\usepackage{lineno}
\usepackage{rotating}
\usepackage{amssymb}
\usepackage{amsmath}
\usepackage{array} 
\usepackage[normalem]{ulem}

\usepackage[colorlinks=true,linkcolor=blue,citecolor=blue]{hyperref}
\newboolean{inbibliography}
\setboolean{inbibliography}{false} 

\begin{document}
\pagenumbering{arabic} %

\begin{titlepage}


{\bf\boldmath\huge
\begin{center}
Extraction of the strong coupling with HERA and EIC inclusive data
\end{center}
}

\begin{center}
Salim Cerci$^{1}$,
Zuhal Seyma Demiroglu$^{2,3}$,
Abhay Deshpande$^{2,3,4}$,
Paul R. Newman$^{5}$,
Barak Schmookler$^{6}$,
Deniz Sunar Cerci$^{1}$,
Katarzyna Wichmann$^{7}$
\bigskip
{\it\footnotesize

$^{1}$ Adiyaman University, Faculty of Arts and Sciences, Department of Physics, Turkiye\\
$^{2}$ Center for Frontiers in Nuclear Science, Stony Brook University, NY 11764, USA\\
$^{3}$ Stony Brook University, Stony Brook, NY 11794-3800, USA\\
$^{4}$ Brookhaven National Laboratory, Upton, NY 11973-5000, USA\\
$^{5}$ School of Physics and Astronomy, University of Birmingham, UK\\
$^{6}$ University of California, Riverside, Department of Physics and Astronomy, CA 92521, USA\\
$^{7}$ Deutsches Elektronen--Synchrotron DESY, Germany\\
}
\end{center}

\vspace{\fill}

\begin{abstract}
\noindent
Sensitivity to the strong coupling $\alpha_S(M^2_Z)$ is investigated
using existing Deep
Inelastic Scattering data from HERA in combination with projected future measurements
from the Electron Ion Collider (EIC) in a next-to-next-to-leading order QCD analysis. 
A potentially world-leading level of precision is achievable when combining
simulated inclusive neutral current EIC data with 
inclusive charged and neutral current measurements from
HERA, with or without the addition of HERA inclusive jet and dijet data.
The result can be obtained
with substantially less than one year of projected EIC
data at the lower end of the EIC centre-of-mass energy range.
Some questions remain over the magnitude of 
uncertainties due to missing
higher orders in the theoretical framework. 
\end{abstract}

\vspace{\fill}

\end{titlepage}



\section{Introduction}
\noindent
Of the coupling strengths of the fundamental forces, the strong coupling $\alpha_s$ is 
by far the least well constrained. 
At the same time, it is an essential ingredient of 
Standard Model cross section calculations, 
as well as constraints on new physics and grand unification scenarios \cite{Georgi:1974sy,Dimopoulos:1981yj}.
It has previously been measured in a wide range of processes \cite{Workman:2022ynf,dEnterria:2022hzv}. In
Deep Inelastic Scattering (DIS), recent studies from HERA have shown limited 
sensitivity when using only inclusive data \cite{H1:2015ubc}, but much more 
competitive precision when additionally including jet production cross 
sections \cite{H1:2015ubc,H1:2021xxi,H1:2017bml}.
In recent years, the advances in QCD theory from next-to-leading order (NLO) to
next-to-next-to-leading order (NNLO) have resulted in a 
substantial reduction in the uncertainties on $\alpha_s$ extractions
due to missing higher order corrections, usually expressed in terms of a QCD
scale uncertainty, though they remain by far the largest single source of uncertainty in
the best HERA extractions. 

\noindent The Electron Ion Collider (EIC)~\cite{Accardi:2012qut},
currently under preparation at 
Brookhaven National Laboratory in partnership with the Thomas 
Jefferson National Accelerator Facility is expected to begin taking data around 2030. 
The EIC will collide highly polarised electrons 
with highly polarised protons 
 and light/heavy nuclei.
In $ep$ mode, the
expected luminosity is of order $10^{33}-10^{34}$~cm$^{-2}$ s$^{-1}$ 
and the centre-of-mass energy $\sqrt{s}$ will range from 
29 GeV to 141~GeV. 
As part of the extensive program of EIC physics~\cite{AbdulKhalek:2021gbh},
inclusive DIS cross sections will be measured to high precision 
in a phase space region that will be complementary to HERA, 
in particular improving the sensitivity to the large Bjorken-$x$ kinematic region.  
In this work, the expected experimental uncertainty on the strong coupling
at the scale of the $Z$-pole mass $\alpha_s(M^2_Z)$ is estimated when
adding simulated EIC inclusive data to analyses very similar to those performed on HERA data. An earlier study of the impact of inclusive EIC
data on $\alpha_s$ precision can be 
found in~\cite{AbdulKhalek:2021gbh}.

\section{Analysis Method}
\label{sec:analysis}

\subsection{Data Samples}
\label{pseudodata}

The HERA data used in this analysis are the final combined H1 and ZEUS 
inclusive DIS neutral current (NC) and charged current (CC) cross sections~\cite{H1:2015ubc}
and, where appropriate, the H1 and ZEUS inclusive and dijet measurements 
used in a recent study of parton distribution function (PDF) sensitivity at NNLO,
as summarised in Table 1 of~\cite{H1:2021xxi}.
The HERA cross sections correspond to unpolarised beam configurations 
at proton beam energies of 920, 820, 575 and 460 GeV and an electron beam energy of 27.5 GeV. 
The data correspond to an integrated luminosity 
of about 1 fb$^{-1}$ and span six orders of magnitude 
in the modules of the four-momentum-transfer squared, $Q^2$, and in Bjorken $x$.

The detailed experimental apparatus configurations for the EIC are currently under
intense development. However, the broad requirements are well established, as 
documented for example in~\cite{AbdulKhalek:2021gbh}. 
In this paper, the simulated EIC data are taken from the studies performed in the
framework for ATHENA detector proposal~\cite{ATHENA:2022hxb}.
The ATHENA configuration has since been combined with ECCE~\cite{adkins2023design}
in the framework
of a new and fast-evolving ePIC design. Whilst the details of the apparatus are
different, the overall kinematic range and achievable precision are expected to be
very similar. 

\noindent As summarised in Table~\ref{tab:samples},
neutral current EIC simulated measurements (`pseudodata') are produced 
with integrated luminosities corresponding to expectations for one year of
data taking with each of the five different beam configurations expected at the
EIC. Charged current pseudodata are also available at the highest $\sqrt{s}$.
The neutral current pseudodata are produced in a grid of five logarithmically-spaced 
$x$ and $Q^2$ values per decade 
over the 
range\footnote{Here, $y$ is the 
usual inelasticity variable, $y = Q^2/(sx)$.}   
$0.001 < y  < 0.95$,
which is well-justified by the expected resolutions. 
The central values are 
taken from predictions using 
HERAPDF2.0NNLO\cite{H1:2015ubc},\footnote{The 
variant with $\alpha_S(M^2_Z)$  set to 0.116 is taken, 
most-closely matching the recent HERA NNLO estimation of $0.1156$~\cite{H1:2015ubc}.}
randomly smeared based on 
Gaussian distributions with standard deviations given by the 
projected uncertainties as estimated by
the ATHENA 
collaboration
and
as previously used to study collinear PDF sensitivities in~\cite{ATHENA:2022hxb,PRN:DIS22}.
The systematic precision is based on experience from HERA and further considerations 
in~\cite{AbdulKhalek:2021gbh} and is rather conservative
in the context of the more modern detector technologies and larger data sets at the EIC. 
Most data points have a point-to-point uncorrelated systematic uncertainty of 
1.9\%, 
extending to 2.75\% at the lowest 
$y$ values. An additional normalisation 
uncertainty of 3.4\% is ascribed, which is taken to be fully correlated between data at each $\sqrt{s}$, 
and fully uncorrelated between data sets with different $\sqrt{s}$. 
For the purposes of the QCD fits (section~\ref{results}), 
the point-to-point systematic uncertainties are added in quadrature with the 
statistical uncertainties and the normalisation uncertainties
are treated as nuisance parameters, as in~\cite{H1:2015ubc}.

\begin{table}[htb!]
\centerline{
\begin{tabular}{|c|c|c|c|}
\hline
$e$-beam energy (GeV) & $p$-beam energy (GeV) & $\sqrt{s}$ (GeV) & Integrated lumi (fb$^{-1}$) \\ \hline 
18 & 275 & 141 & 15.4 \\
10 & 275 & 105 & 100 \\
10 & 100 &  63 & 79.0 \\
5 &  100 &  45 & 61.0 \\
5 &   41 &  29 &  4.4 \\ \hline
\end{tabular}}
\vskip 0.4cm
\caption{Beam energies, centre-of-mass energies and integrated luminosities of the 
different configurations considered for the EIC.
}
\label{tab:samples}
\end{table}

\begin{figure}[htb!]
\begin{center}
\includegraphics[width=0.95\textwidth]{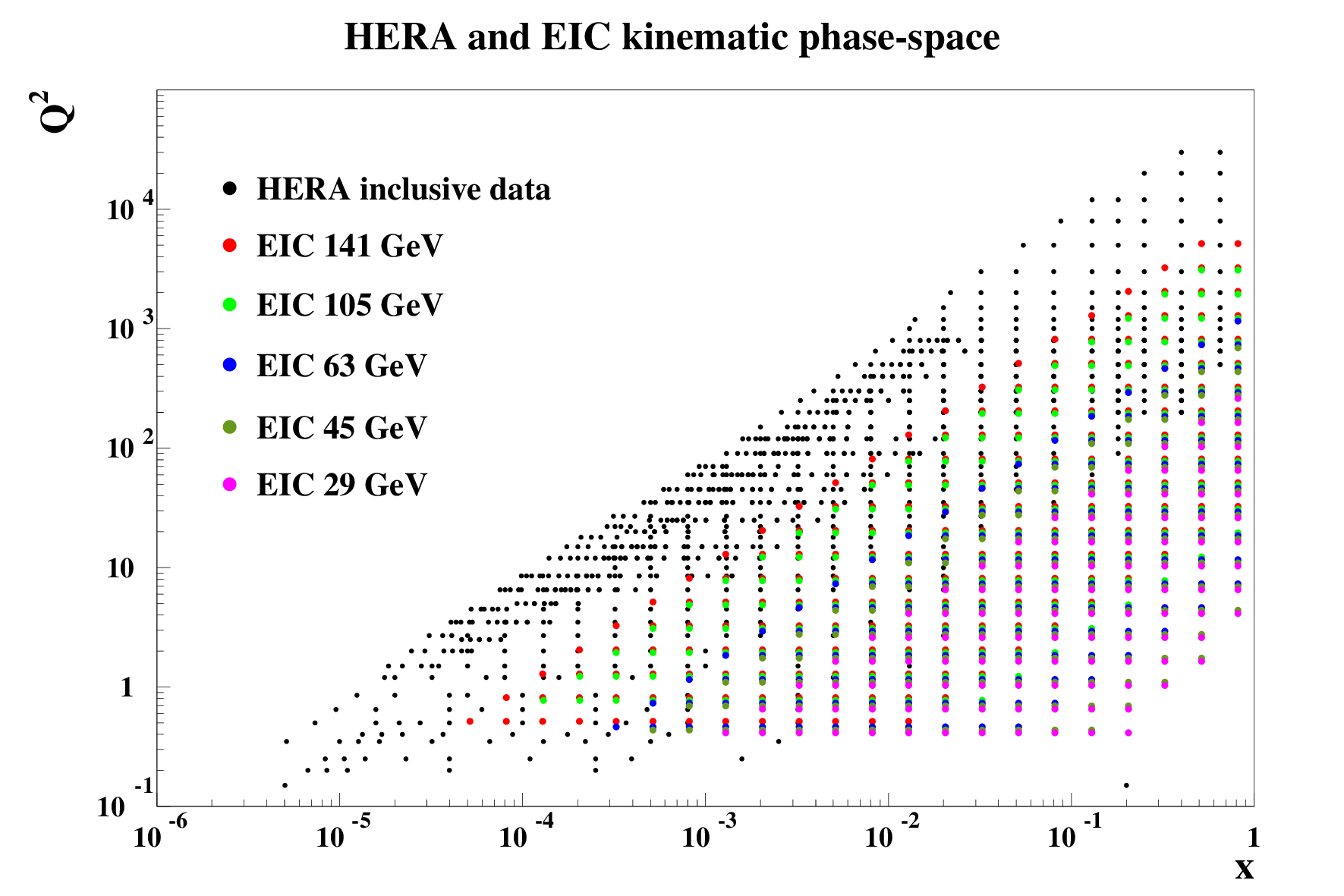}
\caption{The locations in the ($x, Q^2$) plane of the HERA and EIC neutral
current inclusive DIS data points included in the analysis.  
}
\label{fig:phase-space}
\end{center}
\end{figure}

\noindent The kinematic plane of the inclusive data used in this analysis is shown in Fig.~\ref{fig:phase-space}. 
The EIC pseudodata overlap in their coverage with the 
HERA data and extend the kinematic reach in the high $x$, moderate $Q^2$ region. 
Their impact at large $x$ is significant since the large $x$ HERA data are relatively
imprecise due to their kinematic correlation with large $Q^2$, the $1/Q^4$ 
photon propagator term in the cross section and the limited integrated luminosity.

\subsection{Theoretical Framework and Fitting Procedure}

\label{sec:theory}

\noindent The analysis is based on a QCD fit that follows the HERAPDF~\cite{H1:2015ubc} 
theoretical framework, PDF parameterisations and model parameter choices.
In the fit, the proton PDFs and $\alpha_S(M^2_Z)$
are constrained simultaneously in a $\chi^2$-minimisation procedure in which the 
$Q^2$ evolution is performed according to the NNLO DGLAP evolution 
equations~\cite{Gribov:1972ri,Gribov:1972rt,Lipatov:1974qm,Dokshitzer:1977sg,Altarelli:1977zs,Moch:2004pa,Vogt:2004mw,Almasy:2011eq,Mitov:2006ic,Blumlein:2021enk}. 
The xFitter framework~\cite{Alekhin:2014irh,H1:2009pze,H1:2009bcq} is used, with 
the light quark coefficient functions calculated to NNLO as implemented in {\ttfamily QCDNUM}~\cite{Botje:2010ay}. 
The {\ttfamily MINUIT} program~\cite{James:1975dr} is used for the minimisation. 

\noindent
The general-mass variable-flavor-number scheme~\cite{Thorne:1997ga,Thorne:2006qt}
is used for the contributions of heavy quarks. 
The renormalisation and factorisation scales
are taken to be $\mu_r=\mu_f=\sqrt{Q^2}$ for the inclusive DIS data, while 
$\mu^2_r=\mu^2_f=Q^2 + p^2_T $ is used for inclusive jet data and
$\mu^2_r=\mu^2_f=Q^2 + <p_T>^2_2 $ for dijets, where
$<p_T>_2 $ is the average of the transverse momenta of the two jets.
The charm and beauty quark masses ($M_c$, $M_b$) 
follow the choices in~\cite{H1:2015ubc}. 
The minimum $Q^2$ of the inclusive data included in the fits is 
$Q^2_{\rm min} = 3.5$~GeV$^2$. As well as 
avoiding complications associated with low $Q^2$,  
this requirement also reduces the possible influence of $\ln(1/x)$ resummation~\cite{Ball:2017otu}.
An additional cut is applied on the squared hadronic final state invariant mass, 
$W^2 = Q^2 (1-x) / x > 10$~GeV$^2$, which 
removes data points with low $Q^2$ and high $x$ 
that are most likely to be influenced by 
power-like higher twist or resummation effects. 
This cut influences the EIC data sets at the lowest
$\sqrt{s}$. For the central fit, the PDFs are parameterised 
at a starting scale for QCD evolution
of $\mu_{f0} = 1.9$~GeV$^2$, as 
in the HERAPDF2.0 fits~\cite{H1:2015ubc}. 

The PDFs are 
parameterised at the starting scale in terms 
of the gluon distribution ($xg$), 
the valence quark distributions ($xu_v$, $xd_v$), and 
the $u$-type and $d$-type anti-quark distributions
($x\bar{U}$, $x\bar{D}$), 
where $x\bar{U} = x\bar{u}$ corresponds to anti-up
quarks only
and $x\bar{D} = x\bar{d} + x\bar{s}$ is the sum of anti-down and anti-strange
quarks. Symmetry is assumed between the sea quarks and antiquarks for each flavour. 
Strange quarks are suppressed relative to light quarks through a factor $f_s = 0.4$ 
whereby $x\bar{s} = f_s x\bar{D}$ for all $x$.
The nominal parameterisation is
\begin{eqnarray}
\label{eqn:xgpar}
xg(x) &=   & A_g x^{B_g} (1-x)^{C_g} - A_g' x^{B_g'} (1-x)^{25}  ;  \\ 
\label{eqn:xuvpar}
xu_v(x) &=  & A_{u_v} x^{B_{u_v}}  (1-x)^{C_{u_v}}\left(1+E_{u_v}x^2 \right) ; \\ 
\label{eqn:xdvpar}
xd_v(x) &=  & A_{d_v} x^{B_{d_v}}  (1-x)^{C_{d_v}} ; \\ 
\label{eqn:xubarpar}
x\bar{U}(x) &=  & A_{\bar{U}} x^{B_{\bar{U}}} (1-x)^{C_{\bar{U}}}\left(1+D_{\bar{U}}x\right) ; \\ 
\label{eqn:xdbarpar}
x\bar{D}(x) &= & A_{\bar{D}} x^{B_{\bar{D}}} (1-x)^{C_{\bar{D}}} . 
\end{eqnarray}
The parameters $A_{u_v}$ and $A_{d_v}$ are fixed using 
the quark counting rules and $A_g$ 
is fixed using the momentum sum rule. 
The requirement $x\bar{u} = x\bar{d}$ is imposed
as $x \rightarrow 0$ through corresponding
conditions on $A_{\bar{U}}$, $A_{\bar{D}}$,
$B_{\bar{U}}$, $B_{\bar{D}}$ and $f_s$.  
There are thus a total of 14 PDF free parameters. 

The experimental, model, and parameterisation uncertainties on 
$\alpha_s(M^2_Z)$ are evaluated as described in ~\cite{H1:2021xxi}. 
The modelling uncertainties are obtained through variations of $Q^2_{\rm min}$, $f_s$,
$M_c$ and $M_b$ as shown in Table~\ref{tab:model}; the parameters are altered 
independently and the changes relative to the central value 
of $\alpha_s(M^2_{Z})$ are added in quadrature.
For the PDF parameterisation uncertainties, 
the procedure of \cite{H1:2021xxi} is followed, 
based on variations resulting from the
addition of further 
$D$ and $E$ 
parameters to the expressions in Eqs.~\ref{eqn:xgpar}~--~\ref{eqn:xdbarpar}   
and changes in the starting scale 
$\mu^2_{f0}$ by $\pm 0.3$ GeV$^2$.
The fits are repeated with each of these variations  
and the largest difference relative to the nominal $\alpha_s(M^2_{Z})$ is taken to be the uncertainty.
The model and parameterisation uncertainties are added in quadrature in quoting the
final results. 
For the jet data, the
uncertainties in the hadronisation corrections
are treated in the same manner as the experimental  correlated 
systematic uncertainties.

\begin{table}[htb!]
\centerline{
\begin{tabular}{|ll|c|c|c|}
\hline
\multicolumn{2}{|c|}{Parameter} &
\multicolumn{1}{ c|}{Central value} &
\multicolumn{1}{ c|}{Downwards variation} &
\multicolumn{1}{ c|}{Upwards variation}  \\
\hline
$Q^2_{\rm min}$ & [GeV$^2$]& $3.5\phantom{0}$& $2.5\phantom{0^*}$ &
                                          $5.0\phantom{0^*}$   \\
\hline
$f_s$  &  & $0.4\phantom{0}$  & $0.3\phantom{0^*}$ & $0.5\phantom{0^*}$  \\
\hline
$M_c$ & [GeV]  & $1.41$  & $1.37^*$ & $1.45\phantom{^*}$   \\
$M_b$ & [GeV]  & $4.20$  & $4.10\phantom{^*}$ & $4.30\phantom{^*}$  \\
\hline
\end{tabular}}
\vskip 0.4cm
\caption{Central values of model input parameters
  and their one-sigma variations. It was not possible to implement the
  variations marked $^*$ because $\mu_{\rm f0} < M_c$ is required,
  see Ref.~\cite{H1:2021xxi}. In these cases,
  the uncertainty on the PDF obtained from the other variation was symmetrised.
}
\label{tab:model}
\end{table}

The influence on the extracted $\alpha_s(M_Z^2)$ of missing orders in the perturbation
series beyond NNLO is estimated via a scale uncertainty, in which the  
the renormalisation $\mu_r$ and factorisation $\mu_f$ scales 
are varied 
up and down by a factor of two.
Combinations are considered in which $\mu_r$ and $\mu_f$ are changed together
or separately and the 
largest resulting positive and negative deviations on $\alpha_s(M_Z^2)$ 
(with the exclusion of the two extreme combinations of the scales) 
is taken as the scale uncertainty. 
As is currently customary in global QCD 
fits,\footnote{Scale variations are typically
applied to all hadronic final state observables, 
including jet data from $ep$ collisions.} 
no scale variations are made in the 
treatment of the inclusive data.  
This topic is further discussed in section~\ref{theoUncert}.


\section{Results}
\label{results}
\subsection{Fits with EIC Inclusive Data and HERA Inclusive and Jet Data}
\label{res0}

\begin{figure}[htb!]
\begin{center}
\hspace{3mm}\includegraphics[width=0.755\textwidth]{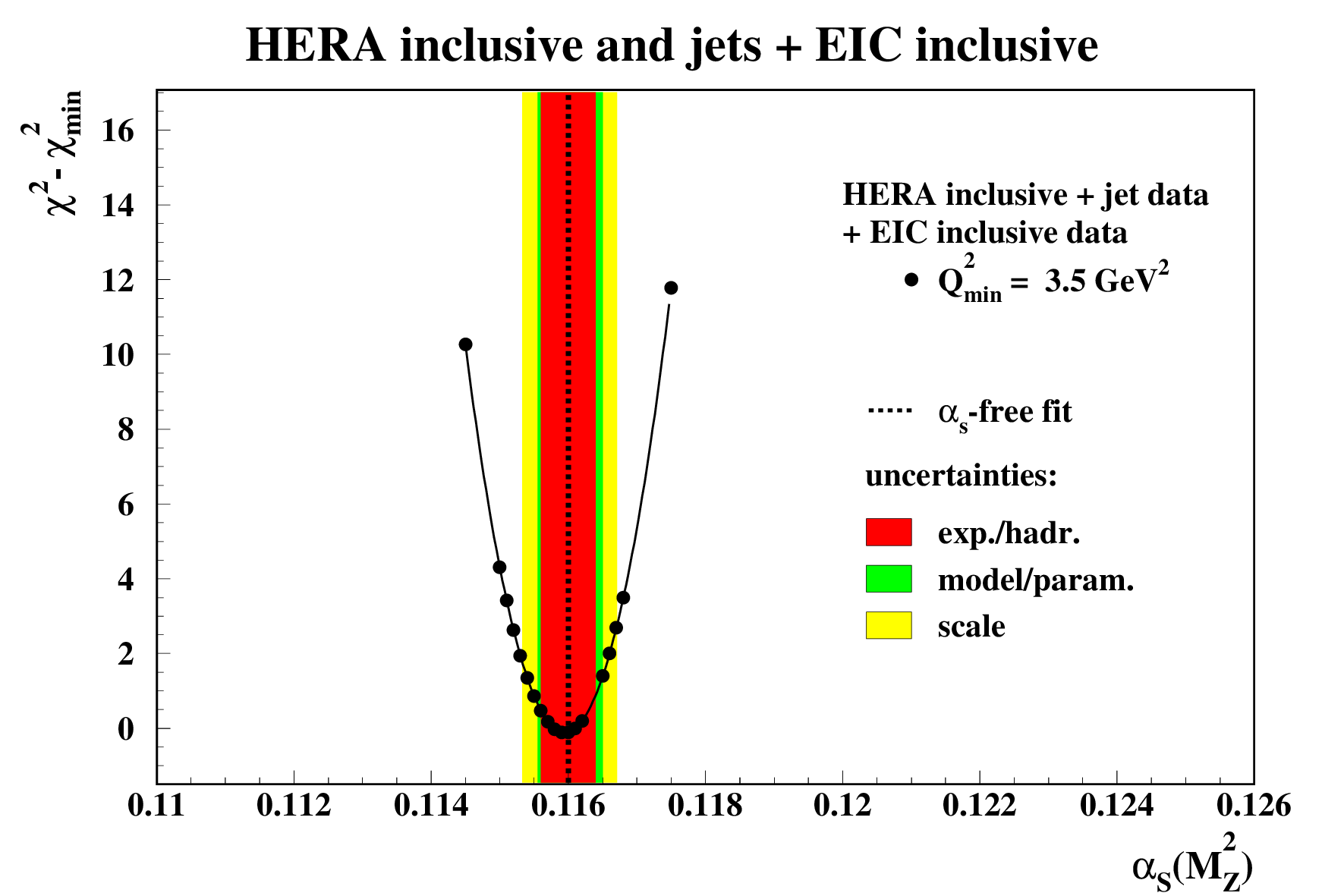} \\
\includegraphics[width=0.715\textwidth]{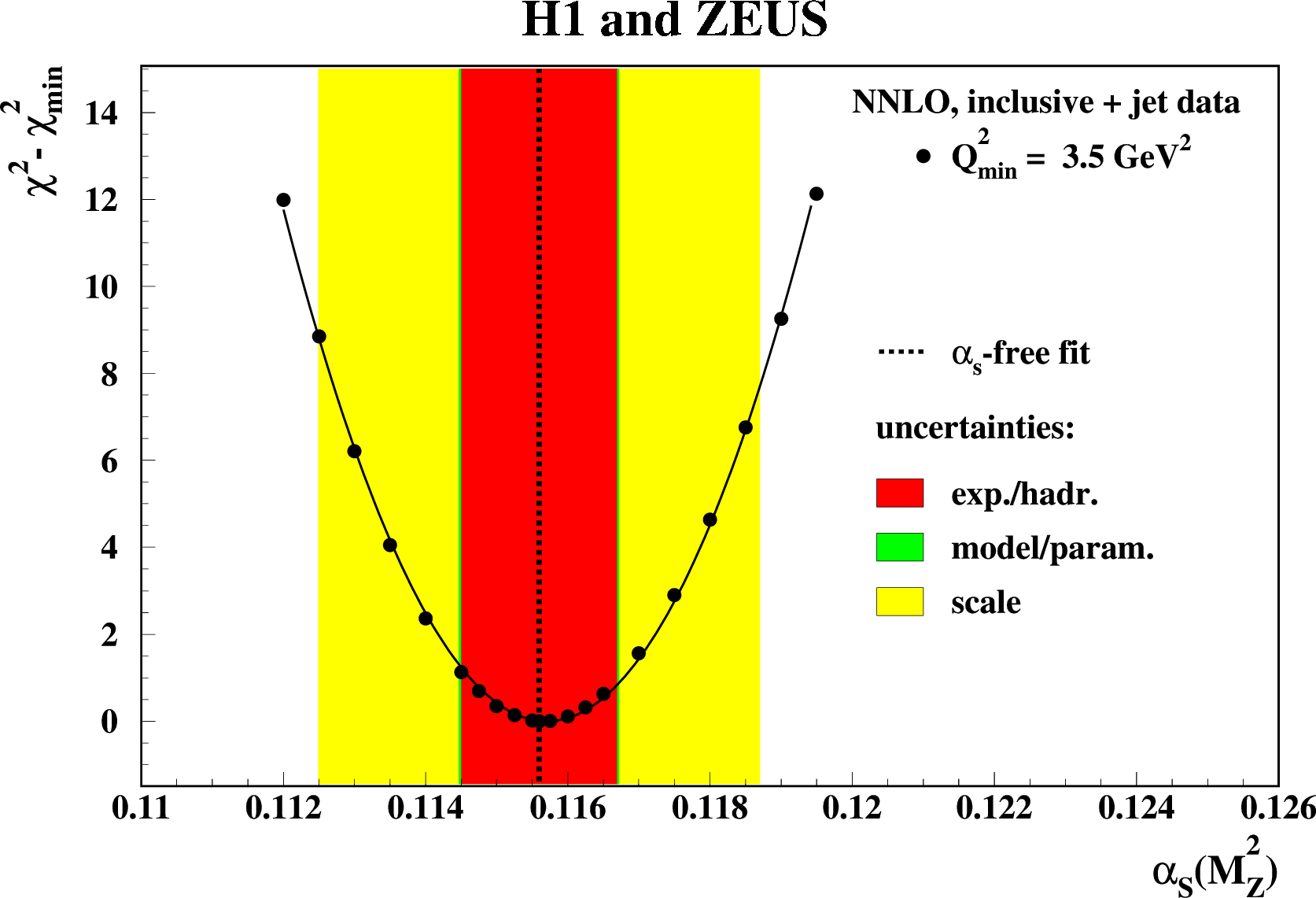} 
\caption{$\Delta\chi^2 = \chi^2 - \chi^2_{min}$ vs. $\alpha_s(M^2_{Z})$ 
for the NNLO fits to HERA and jets data in addition to the simulated EIC inclusive data (top)
and without the EIC data as 
published in~\cite{H1:2021xxi} (bottom). 
The experimental, model, parameterisation, and scale uncertainties are displayed.
}
\label{fig:asJets}
\end{center}
\end{figure}

\noindent 
A simultaneous NNLO fit to extract the PDFs and $\alpha_s(M^2_{Z})$ 
from HERA inclusive and jet data and EIC simulated inclusive data at 
all five $\sqrt{s}$ values is performed as described in 
section~\ref{sec:analysis}. 
The result is
\begin{align}
\alpha_s(M_Z^2) = 0.1160 \pm 0.0004~(\mathrm{exp})~^{+0.0003}_{-0.0002}~(\mathrm{model+parameterisation}) \nonumber
\pm 0.0005~(\mathrm{scale}). 
\end{align}
By construction of the EIC simulated data, $\alpha_s(M_Z^2)$ must be close to 0.116. As expected, the PDF parameters obtained in the fits are also 
fully compatible with those from the HERAPDF2.0 set.
The uncertainties from the joint fit to HERA and EIC data can be compared with those from the HERAPDF2.0Jets NNLO result~\cite{H1:2021xxi}: 
\begin{align}
\alpha_s(M_Z^2) = 0.1156 \pm 0.0011~(\mathrm{exp})~^{+0.0001}_{-0.0002}~(\mathrm{model+parameterisation}) \nonumber
\pm 0.0029~(\mathrm{scale}). 
\end{align}
The results and uncertainties with and without the inclusion of EIC data are shown in
the form of a $\chi^2$ scan as a function of $\alpha_s(M_Z^2)$ 
in Fig.~\ref{fig:asJets}.
Each point in the figure corresponds to a full QCD fit, with all 14 PDF parameters free and a fixed
strong coupling value.
The result without EIC data corresponds exactly to the most recent HERA result~\cite{H1:2021xxi}.

Adding the simulated inclusive EIC data leads to a remarkable improvement 
in the estimated experimental and scale uncertainties. 
The source of the improvement in experimental precision is discussed
in section~\ref{origin}.
The scale uncertainty is reduced to a similar level to the combined model 
and parameterisation uncertainties and becomes smaller than the experimental uncertainty. 
This is a consequence of the reduced dependence of the fit on
the jet data. The scale uncertainty is not yet evaluated for the inclusive
data, as further discussed in section~\ref{theoUncert}.

\subsection{Fits with EIC and HERA Inclusive Data Only}
\label{as}

The very significant impact of the EIC inclusive data on the $\alpha_s(M_Z^2)$ precision
naturally raises the question of whether a similar result can be obtained without the
HERA jet data, i.e. using only inclusive DIS measurements. 
A further question of interest
is how important a role is played by the multiple $\sqrt{s}$ values available at the
EIC. Correspondingly, further fits are performed
to the following inclusive data sets with the fit procedures otherwise unchanged: 
 \begin{itemize}
 \item HERA inclusive data only, as already published in the HERAPDF2 paper~\cite{H1:2015ubc};
 \item HERA inclusive data and the EIC simulated inclusive data described in~\ref{pseudodata}, 
 including all five different $\sqrt{s}$ values in Table~\ref{tab:samples};
 \item HERA inclusive data and the EIC simulated inclusive data, separately for each of the five $\sqrt{s}$ values.
 \end{itemize}

\begin{figure}[htb!]
\begin{center}
\includegraphics[width=0.95\textwidth]{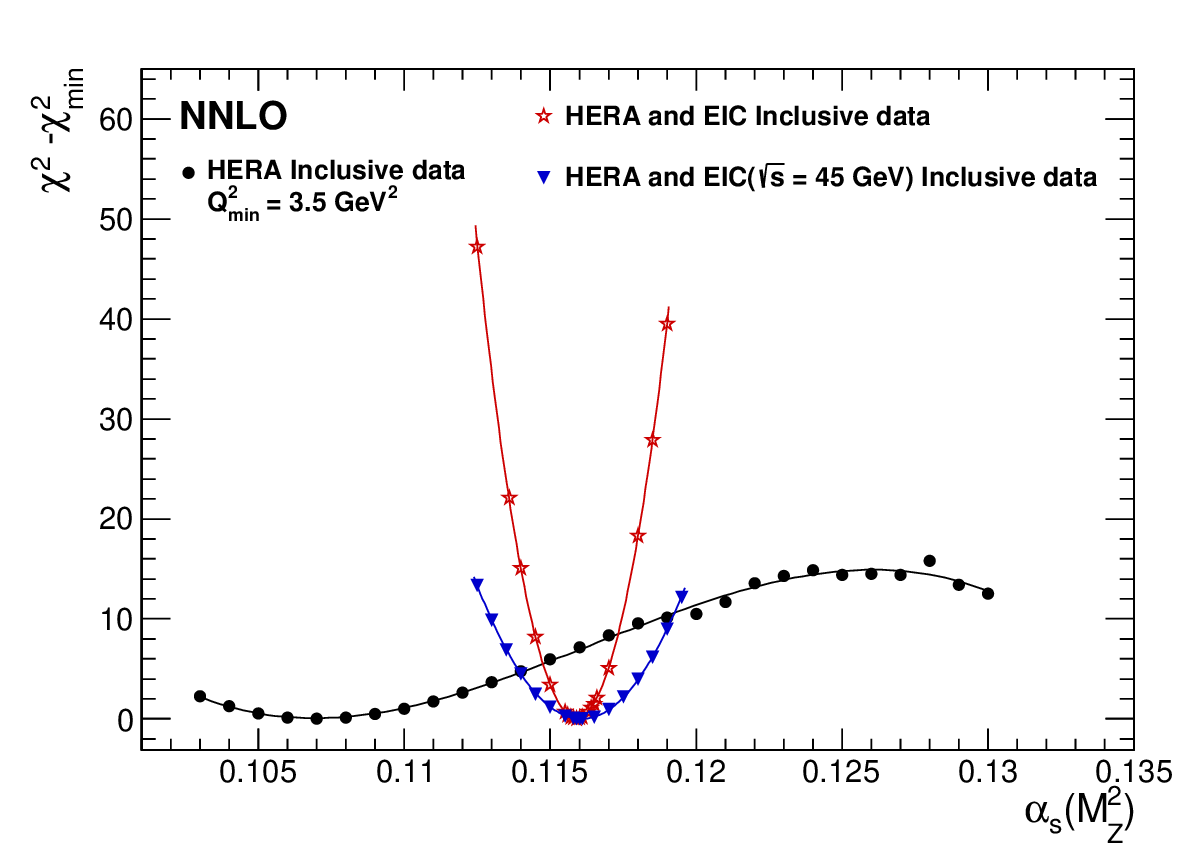}
\caption{$\Delta\chi^2 = \chi^2 - \chi^2_{min}$ vs. $\alpha_s(M^2_{Z})$ for 
the NNLO fits to HERA data on inclusive $ep$ scattering only (black), 
and also with the addition of simulated EIC inclusive data for all five $\sqrt{s}$ values together (red) 
or for only $\sqrt{s} = 45$ GeV (blue). 
The black full points are taken from~\cite{H1:2015ubc}.}
\label{fig:incl1}
\end{center}
\end{figure}

\noindent Figure~\ref{fig:incl1} shows the results of this investigation.
The fits to HERA data alone show only a limited dependence of the fit
$\chi^2$ on the strong coupling $\alpha_s(M^2_Z)$, corresponding to a relatively poor constraint~\cite{H1:2015ubc}. 
In contrast, the $\chi^2$ minimum around $\alpha_s(M^2_Z) = 0.116$ is 
very well pronounced for fits that additionally include EIC data.
Although the best result is obtained when including all EIC $\sqrt{s}$ values
together, 
the precision degrades only slightly when restricting the EIC data to only one EIC $\sqrt{s}$ value. 
In the latter case, the precision improves as the $\sqrt{s}$ value of the chosen 
EIC data decreases. The second lowest $\sqrt{s}$ value, corresponding to 
$E_e \times E_p = 5 \times 100$~GeV, is shown in Fig.~\ref{fig:incl1}. 

The strong coupling extracted from the simultaneous fit for the PDFs and 
$\alpha_s(M^2_{Z})$, using the full set of EIC pseudodata together with the
HERA inclusive data, is 
\begin{align}
\alpha_s(M_Z^2) = 0.1159 \pm 0.0004~(\mathrm{exp})~^{+0.0002}_{-0.0001}~(\mathrm{model+parameterisation}). 
\end{align}
corresponding to a total precision of better than 
0.4\%. 
As discussed in section~\ref{sec:theory}, no scale uncertainty is 
quoted here. It is expected to be significantly
reduced in a fit to inclusive data only relative to the result quoted in 
section~\ref{res0}.
Section~\ref{theoUncert} contains
a discussion of possible ways of estimating the scale 
uncertainties in this case.

The fit using inclusive data only is further extended to investigate the 
influence of the integrated luminosity of the EIC data on the $\alpha_s(M_Z^2)$ precision. 
The statistical uncertainties of the EIC data are scaled such that the pseudodata
samples at each beam energy correspond to 1~fb$^{-1}$, approximately matching the
integrated luminosity of the HERA data. This results in only a small change 
compared with 
the results shown in Fig.~\ref{fig:incl1}.

The relative importance of the different EIC
beam energy configurations has also been investigated. 
When including only a single EIC data set with $\sqrt{s} = 45$ GeV, the experimental uncertainty 
is approximately $\pm 0.0010$, only slightly more than
a factor of two larger than that obtained when including all EIC $\sqrt{s}$ values and 
significantly better than current results from 
DIS data. 
Given that the earliest EIC data are expected to be at low $\sqrt{s}$,
this result might be 
obtainable after significantly less than one 
year of EIC data taking. 
If only the very lowest energy EIC dataset 
($\sqrt{s} = 29$ GeV) is used, the uncertainty
grows considerably, due to the 
influence of the $W^2$ cut.

To further test the influence of the EIC systematic uncertainty assumptions, 
the fit is repeated with the correlated systematic uncertainties 
increased by a factor of two.
The uncertainty on the extracted $\alpha_s(M^2_{Z})$ is barely influenced.
Conversely, 
if the uncorrelated systematic uncertainties are increased by a factor of two,
the uncertainty on $\alpha_s(M^2_{Z})$ increases to around 1.7\%.
The uncorrelated systematic uncertainties are thus the most closely correlated with the 
precision on $\alpha_s(M^2_Z)$.

\subsection{Variations in Analysis Procedure}
\label{robustness}

The robustness of the extracted $\alpha_s(M^2_Z)$ and PDF results and their uncertainties
is tested by varying the details of the fits in a number of ways. 
The relative sensitivity to $\alpha_s$ of different kinematic regions within the 
simulated EIC data is also investigated.

To check for a possible bias from the data simulation procedure, 
the HERA data were replaced with pseudodata obtained using the same method as 
for the EIC samples. 
The $\alpha_s(M^2_{Z})$  scan using the HERA pseudodata 
alone (Fig.~\ref{fig:incl1}) closely follows that of the real HERA data, 
with no distinct minimum observed.

The established technique for including correlated systematic uncertainties in 
global QCD fits treats each source of correlated uncertainty separately, whereas
the EIC estimate is in terms of only a single normalisation uncertainty 
for each $\sqrt{s}$, corresponding to the 
sum of all such sources. Studies are therefore conducted 
in which the correlated EIC systematic uncertainty is decomposed 
into the separate sources, following  
Table 10.5 in the EIC 
Yellow Report~\cite{AbdulKhalek:2021gbh}.
The changes to the results are negligible.

The lowest $Q^2$ data are most likely to be influenced by missing higher orders, higher twist effects 
and $\mathrm{ln}(1/x)$ resummation effects~\cite{Ball:2017otu}. 
To check that the precision is not dramatically altered by excluding these
data, the analysis is repeated with the 
$Q^2_{min}$ cut increased from 
3.5~GeV$^2$ to 10~GeV$^2$ or 20~GeV$^2$.
The distinct minima shown in Figures~\ref{fig:asJets} and~\ref{fig:incl1} 
are still observed, with only a small dependence 
(up to 0.2\%) on $Q^2_{min}$.
Excluding the lowest $x$ EIC and HERA data such that the analysis is restricted
to $x > 0.001$ only increases the uncertainty
on the extracted $\alpha_s$ to 0.0005, although precision is lost in
the PDF determinations.
If all data below $x = 0.01$ are excluded,
the precision on $\alpha_s$ remains at a 
similar level, though the PDF determination
becomes over-parameterised, leading to 
instabilities and biases.

The restriction to $W^2 > 10 \ {\rm GeV^2}$ 
applied here is necessary to
avoid theoretical complications associated with higher
twists or resummations. It removes 
data points with the highest $x$ values at low $Q^2$ 
for the EIC data sets with the lowest $\sqrt{s}$ values, and has no influence on the 
largest $\sqrt{s}$ EIC data or the HERA data.
When only the lowest energy EIC data $\sqrt{s} = 29 \ {\rm GeV^2}$ are included in the 
fit, a systematic dependence on the $W^2$ cut is observed, which is diluted when the 
higher $\sqrt{s}$ data are also included. Nonetheless, 
in the fit to the full HERA and EIC inclusive data, the experimental uncertainty increases
from 0.34\% to 0.52\% when the restriction is altered to $W^2 > 15 \ {\rm GeV^2}$. 
This kinematic region is therefore observed to be important to the EIC $\alpha_s$ sensitivity,
motivating a full understanding of the range of validity of the theoretical
framework as $W$ becomes small. 

\subsection{Discussion}

\begin{figure}[htb!]
\begin{center}
\includegraphics[width=0.95\textwidth]{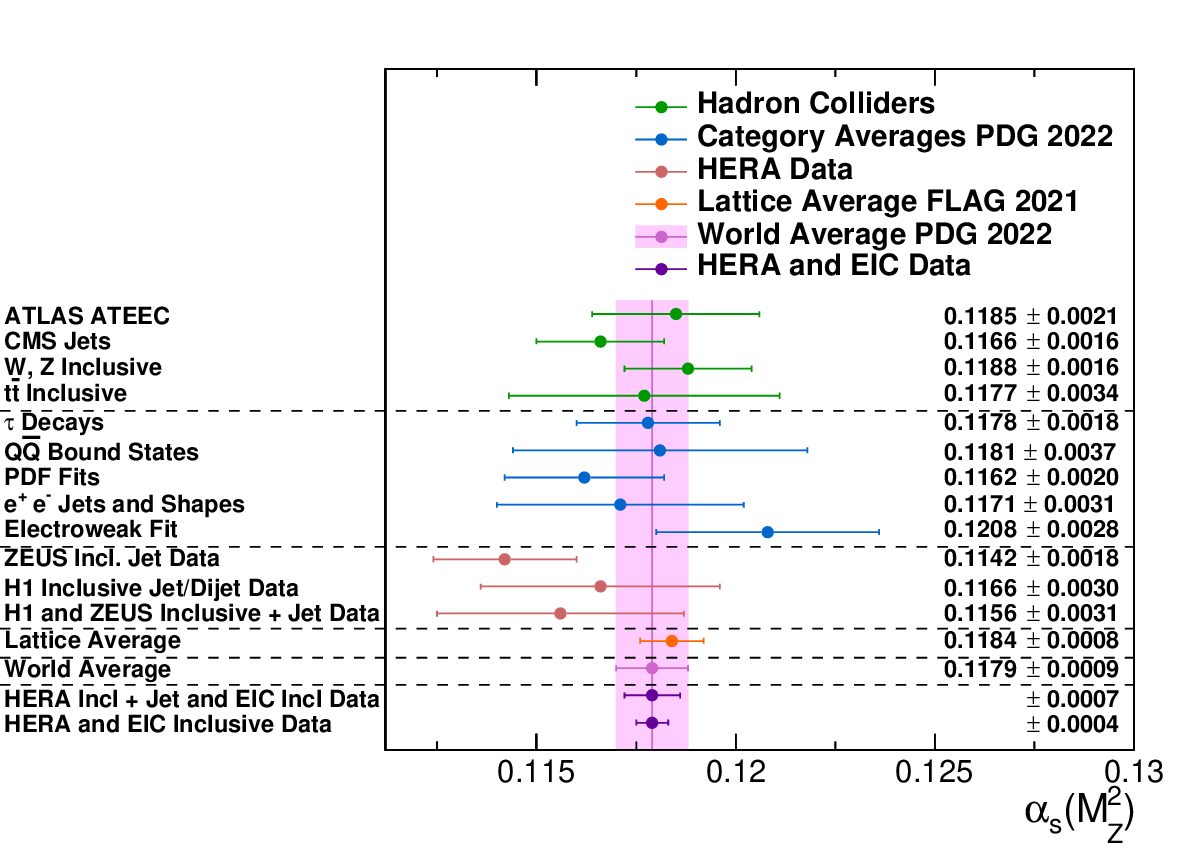}
\caption{Projected total 
uncertainties on the strong coupling constant $\alpha_s(M^2_{Z})$ estimated using HERA and simulated 
EIC data, 
compared with extractions using other data sets and 
methods \cite{ATLAS-CONF-2023-015,CMS:2021yzl,ATLAS:2023tgo,Klijnsma:2017eqp,dEnterria:2019aat,H1:2021xxi,H1:2017bml,Collaboration:2023xha}, 
with the world average 
according to the PDG \cite{Workman:2022ynf}
and with an average from lattice QCD 
calculations \cite{FlavourLatticeAveragingGroupFLAG:2021npn}. Scale uncertainties are not yet included in the
treatment of inclusive DIS data for any of the results shown.  
The plotting style 
follows~\cite{ATLAS-CONF-2023-015}.
}
\label{fig:others}
\end{center}
\end{figure}

The precision on $\alpha_s(M_Z^2)$ obtained in the fits using only inclusive 
HERA and EIC data, and also additionally using HERA jet data,
are compared in Fig.~\ref{fig:others}
with results from previous DIS studies and 
with extractions using a wide
range of other processes. 
The world average of experimental measurements according to the Particle Data
Group (PDG)~\cite{Workman:2022ynf} and an average from lattice QCD 
calculations~\cite{FlavourLatticeAveragingGroupFLAG:2021npn}
are also shown. The projected results from the current 
analyses
show a level of
precision that is significantly better than both the world average
and the lattice average. 
This very encouraging result is subject to the caveat that no
uncertainty has been included 
due to missing higher orders beyond NNLO in the QCD analysis.

\subsubsection{Origin of the EIC Sensitivity}
\label{origin}

The variations in the kinematic range of the
fit described in section~\ref{robustness} show that 
the improvement in experimental precision is attributable to the addition of precise
EIC pseudodata in the large $x$, moderate $Q^2$ region, complementing
the kinematic coverage of the HERA data. This additional phase space coverage leads to improved precision
on the $Q^2$ dependence of the inclusive cross section, 
corresponding to
the logarithmic derivative of the inclusive structure function 
${\rm d} F_2 / {\rm d} \ln Q^2$. 
At the highest $x$ values, this quantity is driven primarily by the
$q \rightarrow qg$ splitting, and therefore samples the product of 
$\alpha_s$ and the large $x$ quark densities. Since $F_2$ is itself a measure of the 
quark densities and other components of
the splitting functions are known
exactly at a given order, 
the logarithmic $Q^2$ derivative 
at large $x$ essentially depends only on
$\alpha_s$.
This contrasts with 
the scaling violations at lower $x$, as well as 
the longitudinal structure function F$_L$~\cite{H1:2013ktq,ZEUS:2014thn} 
and DIS jet data, 
all of which have leading contributions that
are proportional to the product of 
the gluon parton distribution and $\alpha_s$~\cite{Cooper-Sarkar:1987cnv,Zijlstra:1992qd,Boroun:2012bje}.
The improvement in precision can thus be traced to 
the decoupling of $\alpha_s$ from the gluon density, 
enabled by the high $x$ simulated EIC data. This interpretation is supported by
the correlation coefficients between $\alpha_s$ and the other free parameters in the
fit.

\subsubsection{Missing Higher Order Uncertainty}
\label{theoUncert}

Analyses that are sensitive to strong interactions 
commonly
include estimates of the
missing higher order uncertainty (MHOU) in the 
perturbative QCD framework through the 
variation of the renormalisation and factorisation scales
(often referred to as a `scale uncertainty').
As in many other dedicated and global analyses, the approach used for the
jet data included here is to obtain a 
scale uncertainty by varying the scales by factors of two.
However, in the global 
QCD fits to extract 
PDFs \cite{NNPDF:2021njg,Bailey:2020ooq,Hou:2019efy,Alekhin:2017kpj},
MHOUs have routinely not been included in the treatment of
inclusive DIS data, since they are expected to be relatively small
in comparison with other PDF uncertainties.  
Since the present analysis adopts 
a perturbative QCD treatment at NNLO, it is reasonable to assume that 
it also has relatively small
MHOUs, and the 
situation is expected to improve further as
the state-of-the-art moves towards 
N$^3$LO \cite{McGowan:2022nag}. 
Nevertheless, the MHOU 
associated with inclusive data must clearly be finite and cannot be ignored completely
at the very high level of precision suggested in 
the present analysis. First studies of their influence on PDFs have been performed by the NNPDF collaboration~\cite{Ball:2017otu}, 
though the impact on the strong coupling was not included.
There is as yet no consensus how to estimate MHOUs for inclusive DIS.
Some discussion of possible methods is supplied in the following. 

In a previous analysis at NLO accuracy~\cite{H1:2000muc}, the H1 collaboration 
made a combined fit of
inclusive-only DIS data from HERA and from the fixed target BCDMS experiment.
The strong coupling was found to be well-constrained, with the BCDMS data
playing a similar role to the EIC pseuododata here. A MHOU was obtained
by varying the factorisation and renormalisation scales in the  usual way,
resulting in a large uncertainty at the level of 4\%. However, as shown in the context of PDF uncertainties in~\cite{NNPDF:2019ubu} (Appendix B),
applying this method in an NLO analysis results in an estimate of the MHOU
that is larger than the difference between NLO and NNLO results
by a factor as large as $20-50$. 

\noindent
A similarly conservative approach might
be to fit pseudodata simulated using QCD evolution at
NNLO using an NLO framework and vice versa, taking the 
MHOU on $\alpha_s(M^2_Z)$ to be the deviation of the extracted
$\alpha_s(M^2_Z)$ from the input. Applying this method to the present
analysis also results in an uncertainty of around 4\%, but is also
likely to be a very significant over-estimate. 

\noindent 
A potentially promising approach is suggested by the NNPDF group~\cite{NNPDF:2019ubu}.
First a theory covariance matrix is computed, typically 
using scale variations to include 
missing higher order uncertainties.
Including the covariance matrix explicitly 
in the PDF fit ensures that the theory  
uncertainties propagate properly,
including those associated with  
$\alpha_s(M^2_Z)$ if it is a fit parameter. However, until a consensus around a well-developed method 
for including inclusive data such as this 
emerges, the MHOU in the present $\alpha_s(M^2_Z)$
extraction remains to be evaluated.

\section{Conclusions and Outlook}
\noindent This work shows that the strong coupling 
can be determined with potentially world-leading 
precision in a simultaneous fit of 
PDFs and $\alpha_s(M^2_{Z})$ at NNLO in perturbative QCD, 
using only inclusive DIS data from HERA and simulated data from the EIC. 
The estimated uncertainty on the strong coupling 
when including one year's data at each of the five expected EIC
$\sqrt{s}$ values is better than 0.4\%, 
substantially improving on the precision of the 
present world experimental and lattice averages.
If the EIC pseudodata are restricted to a small 
fraction of a standard
expected year of running at a 
centre-of-mass energy of 45 GeV, 
as expected in an early phase of operation, 
the estimated total uncertainty is at the 
level of 0.9\%. 
The improvement in precision is traceable to the large $x$, 
intermediate $Q^2$
region that was not accessed at HERA, but is well covered by the EIC. 
The constraint arises
primarily from the evolution of the 
quark densities in this region and 
is largely decoupled from the uncertainty
on the gluon density. 
It still remains to assign a meaningful uncertainty due to missing higher order contributions beyond NNLO in the theory.

\noindent Further improvements of the $\alpha_s(M^2_{Z})$ precision may be 
obtainable
by adding inclusive jet and dijet EIC data to the QCD analysis,
for example using theory grids for the EIC energies in the fastNLO framework~\cite{Kluge:2006xs}. 
Other observables carrying information on the strong 
coupling that may be measured at the EIC
include event shapes, jet substructure and jet radius parameters.
As well as a DIS-only approach, it would also be 
interesting to investigate the impact of EIC data on
$\alpha_s$ determinations in global QCD 
fits that also include data from the LHC and
elsewhere \cite{NNPDF:2021njg,Bailey:2020ooq,Hou:2019efy,Alekhin:2017kpj}.

\noindent In the time before the start of the EIC, 
it is hoped that new light will be shed on the issue of higher order 
uncertainties,
leading to a consensus on how they should be treated in  
$\alpha_s(M^2_Z)$ determinations relying on EIC data.

\section{Acknowledgements}
We are grateful to many colleagues in the EIC experimental community who have worked on all aspects of
the project over many years and are now developing detector concepts towards the 
acquisition of real data similar to those simulated here. 
We thank numerous theoretical physics colleagues for very valuable discussions about the theory uncertainties:  
N\'{e}stor Armesto, Andrea Barontini, Thomas Cridge, Stefano Forte, Lucian Harland-Lang, Anna M. Sta\'{s}to and Robert S. Thorne,
as well as Valerio Bertone and Francesco Giuli for their help with the {\tt APFEL} program 
and Christopher Schwan for his help with the {\ttfamily PineAPPL} tool. This work of Z.S.D and A.D. was supported in part by the U.S. Department of Energy and the Simons Foundation.

\pagebreak
\setboolean{inbibliography}{true}
\addcontentsline{toc}{section}{References}
\bibliographystyle{JHEP}
\bibliography{references}
\setboolean{inbibliography}{false}

\end{document}